\documentclass[twocolumn,times]{aastex631} %
\usepackage{amsmath}

\newcommand{\Rs}{R_{\odot}}
\newcommand{\kms}{\mathrm{km\,s^{-1}}}

\shorttitle{Model of Birth and Cannibalism of Homologous CMEs}
\shortauthors{Cai et al.}
\submitjournal{ApJL}

\begin{document}

\title{A Global Magnetohydrodynamic Simulation of the Birth and Cannibalism of Homologous Coronal Mass Ejections}

\author{Lijun Cai}
\affiliation{State Key Laboratory of Solar Activity and Space Weather, School of Aerospace, Harbin Institute of Technology, Shenzhen 518055, China}
\author{Chaowei Jiang}
\email{chaowei@hit.edu.cn}
\affiliation{State Key Laboratory of Solar Activity and Space Weather, School of Aerospace, Harbin Institute of Technology, Shenzhen 518055, China}
\author{Zhenhua Liu}
\affiliation{State Key Laboratory of Solar Activity and Space Weather, School of Aerospace, Harbin Institute of Technology, Shenzhen 518055, China}
\author{Xueshang Feng}
\affiliation{State Key Laboratory of Solar Activity and Space Weather, School of Aerospace, Harbin Institute of Technology, Shenzhen 518055, China}

\correspondingauthor{Chaowei Jiang}

\begin{abstract}
Homologous coronal mass ejections (CMEs) are sequential eruptions from the same active region. They are of particular space weather interest because they can interact and coalesce into compound ejecta, potentially amplifying their geomagnetic impact. Yet the full chain from recurrent initiation to magnetic merging has not been captured self-consistently in a single global simulation. We present a three-dimensional magnetohydrodynamic model that generates three homologous CMEs from a bipolar active region by recurrent photospheric shearing. All three eruptions follow the same initiation mechanism by tether-cutting reconnection within a recurrently formed sigmoid, while the later CMEs propagate significantly faster than the first. We demonstrate that this speed hierarchy arises from progressive coronal preconditioning; preceding eruptions evacuate the overlying corona, reducing both mass loading and ambient drag. 
The faster later CMEs catch up with their predecessors, and the compression between them forms inter-CME current sheets. Reconnection within these current sheets then progressively merges the interacting flux systems, completing the cannibalistic process.
Synthetic coronagraph images reproduce the apparent coalescence of multiple white-light fronts. This simulation provides the first self-consistent global view of the birth, preconditioning, and cannibalistic interaction of homologous CMEs, bridging recurrent eruption initiation and inter-CME magnetic merging in the near-Sun corona.
\end{abstract}


\section{Introduction}\label{sec:intro}

Coronal mass ejections (CMEs) are large-scale eruptions of magnetized plasma from the solar corona and are major drivers of space weather. Their geoeffectiveness depends on the speed, trajectory, and magnetic configuration of the ejecta, all of which evolve dynamically from CME birth in the low corona to their transit in the heliosphere. The evolution is further complicated when multiple CMEs erupt within a short interval and interact with one another, which are not unusual during maximum periods of solar activity~\citep{Yashiro2004,Robbrecht2009}. Their interactions could significantly enhance the geoeffectiveness compared with isolated events~\citep{Liu2014,Lugaz2017,Shen2018,Scolini2020,Palmerio2021,soni2023}. For instance, \citet{Liu2014} showed that, in the 2012 July 23 event, an earlier eruption preconditioned the upstream solar wind, contributing to the weak deceleration and short transit of the later event; the subsequent interaction of two closely launched CMEs compressed the ejecta and amplified its magnetic field, resulting in a so-called ``perfect storm.''

Homologous CMEs denote sequential eruptions originating from the same active region (AR), thereby sharing similar propagation directions and magnetic topologies, which are particularly favorable for magnetic interactions between the CMEs. Observations link the initiation of such sequences to sustained shear and continued energy storage \citep{Li2010}, repeated flux emergence, cancellation, and reconnection \citep{Panesar2016}, and recurrent kinked rope-like source structures \citep{Liu2019}. After initiation, later CMEs could cannibalize their preceding ones, forming complex compound ejecta. Recently, \citet{Song2026} reported four homologous CMEs launched from AR 11303 within ten hours that caught up and coalesced in the near-Sun corona. \citet{Soni2026} suggested that the 2024 Mother's Day Event, a historic geomagnetic storm, is produced by interactions of over ten CMEs ejected predominantly from AR 13664 within four days. 

Current numerical magnetohydrodynamic (MHD) studies have produced homologous eruptions through continuous shear and repeated breakout reconnection~\citep{DeVore2008}, continued emergence of twisted magnetic flux rope (MFR) that undergoes repeated partial eruptions due to kink instability~\citep{Chatterjee2013}, and recurrent formation and disruption of a current sheet above the same polarity inversion line (PIL) of a bipolar arcade as driven by continual shearing~\citep{Bian2022}. However, these simulations are {mainly} limited to resolving the eruption onset, leaving the complex interactions of ejecta unexplored. {Although the simulation of \citet{Chatterjee2013} has partially captured the cannibalism, the magnetic interaction during this process has not been studied. Moreover, that simulation was performed within a spherical wedge spanning $15^{\circ}$ in latitude and $30^\circ$ in longitude. Such a limited domain is too small to characterize the development and interaction of CMEs self-consistently, owing  to the strong influence of the lateral boundaries. Global simulations covering the full $4\pi$ sphere are therefore highly desirable.} Global coronal and heliospheric simulations have been used to study subsequent CME--CME interactions by inserting non-equilibrium MFRs into the computational domain at sequentially separated times, sometimes with imposed initial speeds~\citep[e.g.,][]{Lugaz2005, Lugaz2007, Bemporad2012, Shen2013, Shiota2016, Scolini2020, Soni2026}. However, these simulations bypass the initiation phase of the eruptions. The interplay between successive CMEs may set in before the later eruption has fully lifted off, since the preceding CME has already altered the background coronal conditions~\citep{Torok2011}. This argues against decoupling eruption initiation from propagation in a self-consistent simulation.

This Letter presents the first global three-dimensional (3D) MHD simulation that captures self-consistently both the birth and cannibalistic interactions of homologous CMEs from a bipolar-arcade AR, recurrently energized by photospheric shearing motion. Three CMEs are initiated in sequence through the same sheared-arcade tether-cutting reconnection mechanism, yet each successive eruption produces a faster CME as a result of progressive coronal preconditioning by the preceding ejecta. 
The later CMEs consequently catch up with and cannibalize the earlier ones, eventually merging their distinct magnetic systems into a compound ejecta. The simulation further reveals the reconnection processes that govern the magnetic coalescence of the interacting CMEs, which have rarely been elucidated in previous 3D simulations.

\section{Numerical Model}\label{sec:model}

\begin{figure*}
    \centering
    \includegraphics[width=\textwidth]{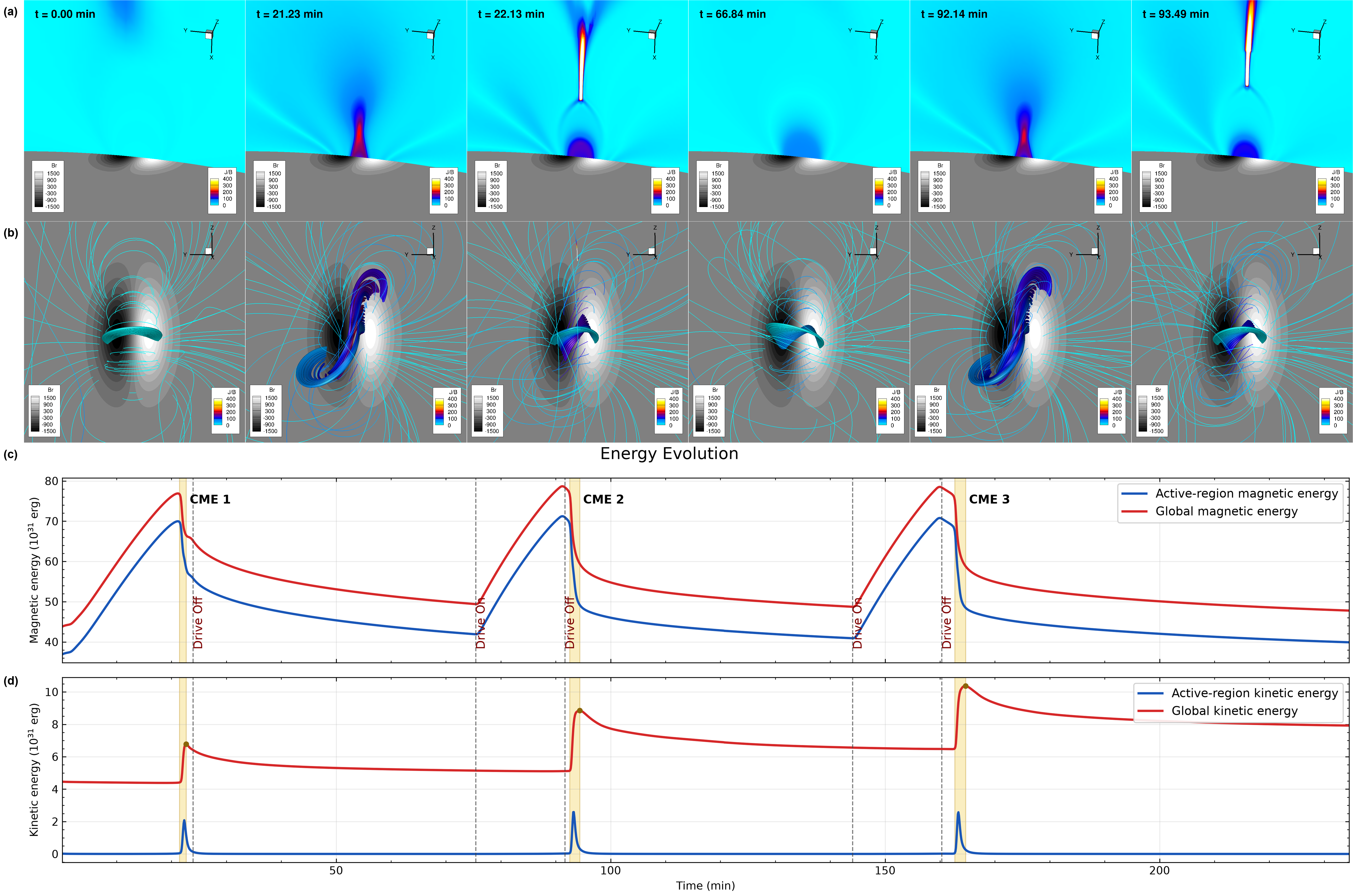}
    \caption{Evolution of energies and source-region topology during three homologous eruptions. Panels (a) and (b) show the repeated arcade--sigmoid--arcade transition, with (a) displaying the normalized current density $J/B$ on a vertical slice and (b) showing sampled 3D field lines above the photospheric $B_r$ distribution. Panels (c) and (d) present the magnetic and kinetic energy evolution for the global and AR volumes, respectively. Vertical dashed lines mark each energizing episode; yellow shading indicates the impulsive acceleration phase of each CME. (An animation of this figure is available. {It displays the evolution of panels (a) and (b) over the entire simulation period from $t=0$ to $t=233$~min with a cadence of $0.45$~min.})
    }
\label{fig:energy}
\end{figure*}

\begin{figure*}
    \centering
    \includegraphics[width=\textwidth]{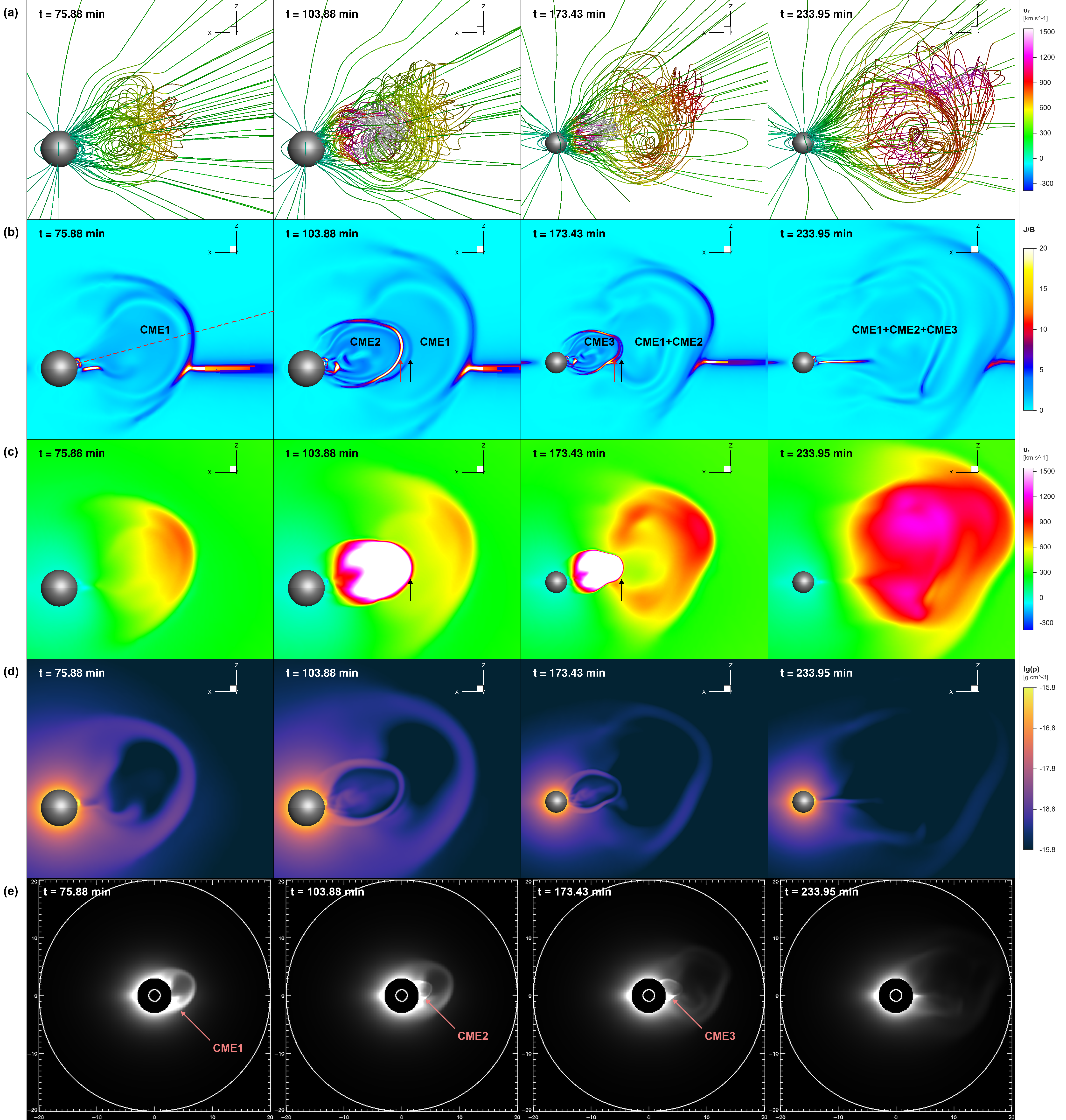}
    \caption{Global evolution toward compound ejecta and synthetic coronagraph signatures. (a)--(d) Snapshots at four selected times, illustrating the formation and interaction of three CMEs, with (a) displaying 3D magnetic field lines color-coded by radial velocity, (b) the normalized current density $J/B$ on the central meridional plane, (c) the radial velocity, and (d) the logarithmic density. The sphere represents solar surface. The dashed line in panel (b) marks the sampling path for the time--distance analysis in Figure~\ref{fig:td}. {The two arrows in the second frame (i.e., $t=103.9$~min) of panel (b) mark the MFR front (by red arrow) and the CME-driven shock (black arrow), respectively. In panel (c), the black arrow is also shown denoting the same location in the velocity profile. The arrows in the third frame ($t=173.4$~min) have the same meanings as those in the second frame, but for CME3.} (e) Synthetic polarized-brightness images at four selected times, with a field of view of $20\Rs$ centered on the Sun and an inner occulter at $3\Rs$. {The line of sight is defined along the $y$ axis. The inner circle is the solar surface and the outer circle is $20\Rs$.} (An animation of this figure is available. {It displays the evolution of panels (a)-(d) over the entire simulation period from $t=0$ to $t=233$~min with a cadence of $0.45$~min.})}
    \label{fig:global}
\end{figure*}

We use the global-corona component of the Sun-to-Earth CME evolution MHD modeling framework recently developed by~\citet{Jiang2026}. The model solves a set of modified MHD equations in a Carrington co-rotating frame including solar gravity and Coriolis force. The thermodynamics are simplified as a polytropic process with an effective polytropic index of $\gamma=1.05$. A semi-relativistic Boris correction \citep{Boris1970,Gombosi2002,Matsumoto2019} combined with a relativistic mass-density factor and an artificially reduced light speed of $1.28 \times 10^4~ \kms$ is introduced in the momentum equation to limit the effective Alfv\'en speed in magnetic fields up to thousands of gauss. The MHD equations are solved by the advanced conservation element and solution element in general curvilinear geometry with adaptive mesh refinement~\citep{Jiang2010,Jiang2012}.

The computation uses an overset Yin--Yang spherical grid with a base angular resolution of $2^{\circ}$. Six refinement levels provide a finest angular resolution of $0.0625^{\circ}$, corresponding to approximately $700$~km in the low corona. The refinement follows gradients of the MHD variables and dynamically captures density, momentum, and Lorentz-force structures associated with eruptive structures, and particularly maintains the strong-field AR and the principal current sheets at the highest refinement level. The domain extends from solar surface to $20\Rs$. The lower boundary fixes the density and temperature, evolves the magnetic field with the induction equation by imposing a surface electric field $\mathbf{E}_s$. At the outer boundary, which is assumed to fully open, all MHD variables are extrapolated with zero gradients along the radial direction.

The initial conditions are constructed following~\citet{Jiang2026}. We first relax a Parker's solar wind solution (with density of $5.51 \times 10^{-17}$~g~cm$^{-3}$ at the lower boundary and temperature of $1.8\times10^6$~K) within a global magnetic dipole (with a maximum surface field of $1.5$~G at the poles) to an equilibrium, and then emerge a bipolar AR into the background.  As shown in Figure~\ref{fig:energy}a--b~\citep[and also Figure~1 in][]{Jiang2026}, the AR has scale of around $100$~Mm, maximum field strength of $1000$~G and unsigned magnetic flux of $1.7\times10^{22}$~Mx, which are typical in realistic ARs~\citep{vandriel2015,toriumi2019}. It is centered at Carrington longitude $180^{\circ}$W and latitude $15^{\circ}$N.

To energize the AR, we apply a surface rotational flow $\mathbf{v}_d$ to its two magnetic polarities, and thus the solar surface electric field is given by $\mathbf{E}_s = -\mathbf{v}_d \times \mathbf{B}$. The flow $\mathbf{v}_d$~\citep[defined by Equation~14 in][]{Jiang2026} follows the contours of the radial magnetic field $B_r$ and preserves the magnetic flux distribution while injecting free magnetic energy. For computational efficiency, the flow has an average speed of around $10\,\kms$, much faster than typical photospheric motions. Nevertheless, it remains below the effective local Alfv\'en speed by three orders of magnitude, and still represents a quasi-static driving to the model. 

While \citet{Jiang2026} applied the surface driving flow continuously until the first eruption and then switched it off, we apply it in three episodes to sequentially produce three eruptions, denoted CME1, CME2, and CME3 (see the energy evolution in Figure~\ref{fig:energy}c--d). The surface flow is stopped at closely the onset of each eruption and is applied again after an interval of around $50$ minute. CME2 and CME3 therefore originate from the post-eruption source region and propagate through the disturbances left by preceding events. The interval is selected so that all three CMEs and their near-Sun interactions can be clearly tracked within the $20\Rs$ domain. Test runs show that shorter intervals cause the eruptions to interact too early and merge within only a few solar radii.

\section{Results}\label{sec:results}

\subsection{Energy Evolution and CME Initiation Mechanism}\label{subsec:energy}

Figure~\ref{fig:energy}c-d shows evolution of the magnetic and kinetic energies for both the AR and the global volumes. During each energizing episode, the magnetic energy increases continuously as the surface flow injects magnetic energy into the corona through upward Poynting flux. The AR kinetic energy remains close to zero before each eruption, showing that the system passes through a sequence of quasi-static states. Each eruption is marked by an impulsive increase in kinetic energy and a corresponding rapid decrease in magnetic energy. According to the profile of the global kinetic energy, the CME acceleration phase can be identified by the interval during which the kinetic energy begins to rise abruptly until its peak value, as denoted by the yellow-shaded areas in the figure. This phase takes only a few minutes. The subsequent decline in global kinetic energy reflects deceleration of CME due to its expansion and interaction with the ambient corona or a preceding CME; the later plateau is consistent with the interacting structures beginning to propagate together. After each CME's acceleration phase, the magnetic energy continues to decrease, but at a much slower rate. The AR kinetic-energy curves appear as narrow spikes because the erupting structures rapidly leave the AR sub-volume, causing the local kinetic energy to return quickly to its pre-eruption level.

Immediately before all three eruptions, the AR magnetic energy is approximately the same value of $7\times10^{32}$~erg. This is consistent with previous Cartesian simulations showing that a sheared arcade has an upper energy limit set by the open-field energy and must erupt when that limit is approached~\citep{Jiang2021,Bian2022}. In the CME acceleration phase, the released  magnetic energy is around $1.2\times 10^{32}$~erg in CME1 and around $2.1\times 10^{32}$~erg in both CME2 and CME3; the gained kinetic energy is $2.1\times 10^{31}$~erg in CME1 and around $4\times 10^{31}$~erg in both CME2 and CME3. Thus, the later two eruptions release significantly more magnetic energy and produce stronger kinetic responses than the first one. This indicates that the CME2 and CME3 are faster than CME1 and CME--CME collisions could occur.

Figure~\ref{fig:energy}a--b (and the accompanying animation) shows evolution of magnetic configuration and current density distribution in the AR. The surface flow progressively transforms the initially potential-field arcade ($t=0$~min) into a strongly-sheared S-shaped arcade (i.e., sigmoid) before the first eruption ($t=21.2$~min). A thin, intense current sheet develops within the core field above the PIL, and reconnection in that current sheet triggers and drives the eruption, a fundamental mechanism of CME initiation first conjectured by~\citet{Moore1992} and demonstrated by~\citet{Jiang2021}. Shortly after CME1 onset ($t=22.1$~min), reconnection has opened and reconfigured part of the erupting field, while the post-eruption arcade retains appreciable shear. During the interval with the surface driving flow switched off, that residual nonpotential structure relaxes gradually ($t=66.8$~min). The subsequent shearing episodes begin from arcade that already contains free energy, and less additional photospheric energy input is required to reach the next eruptive state than the first eruption. The following two eruptions share the same initiation mechanism of current-sheet formation and subsequent reconnection. The full simulation reproduces the recurrent arcade--sigmoid--arcade transition of magnetic topology in source regions of typical observed homologous CMEs.

\subsection{Global Evolution toward Compound Ejecta}\label{subsec:global}

Figure~\ref{fig:global}a--d (and the accompanying animation) show evolutions of 3D magnetic field lines and a central cross section (i.e., the $y=0$ plane) of current density normalized by magnetic field strength ($J/B$), radial velocity, and the plasma density, depicting formation of the CMEs and their interactions. At $t=75.9$~minutes CME1 is the only ejecta in the outer corona. At $103.9$~minutes CME2 has erupted and is catching up with CME1, while the magnetic flux systems of the two CMEs remain distinguishable. At $177.0$~minutes CME1 and CME2 form a compound structure while CME3 has erupted and begins to interact with the preceding ejecta. By $234.0$~minutes the main magnetic fluxes of the three CMEs are largely merged.

CME1 has among the most regular morphology, with an approximately symmetric expansion and a relatively coherent density envelope. CME2 and CME3 are more radially stretched and asymmetric because they propagate through wakes whose density, velocity, and magnetic structure have been changed by preceding eruption. The heliospheric current sheet, which initially occupies the $z=0$ plane, is distinct before eruption, becomes strongly distorted or indistinguishable inside the CME volume, and gradually reforms behind the ejecta.

{The distribution of $J/B$ shows enhanced fronts for all the CMEs. For CME1, there is a single front that corresponds to the shock, across which the radial velocity changes abruptly (compare Figure~\ref{fig:global}b and c). In contrast, for CME2 and CME3, the initially single front evolves into two distinct fronts at the CME nose. Their locations are marked by the two arrows in the frame at $t=103.9$~min for CME2 and by the arrows in the frame at $t=173.4$~min for CME3.} The outer front is the CME-driven shock, which propagates through the preceding CME MFR, where the Alfv\'en speed is higher than in the unperturbed background, allowing the later shock to be transmitted rapidly. {The inner front, by contrast, corresponds to the interface between the MFR of the new CME and that of the preceding CME, which is to be examined in Section~\ref{subsec:kinematics}.} 

Figure~\ref{fig:global}e presents synthetic polarized brightness from line-of-sight Thomson scattering, designed to mimic coronagraph observations. {Initially, all the CMEs exhibit a bright front owing to the density compression ahead of their MFRs, along with a relatively dark cavity resulting from the density decrease caused by the rapid expansion of these MFRs. At $103.9$ minutes, the front of CME2 appears within the wake of CME1. By $173.4$ minutes, CME1 and CME2 have merged into a single front, while the front of CME3 has entered into the cavity of the preceding CMEs.} At $233.9$ minutes, the three CMEs appear to merge into a broad, asymmetric compound signature, in which individual CME fronts are difficult to identify from brightness alone. The synthetic images demonstrate how several magnetically distinct ejecta can appear to coalesce in coronagraph data. They do not, by themselves, establish magnetic merging; that inference requires the current-density and field-line diagnostics in Section~\ref{subsec:kinematics}.

\subsection{CME Kinematics, Interaction, and Magnetic Merging}\label{subsec:kinematics}

\begin{figure*}
    \centering
    \includegraphics[width=\textwidth]{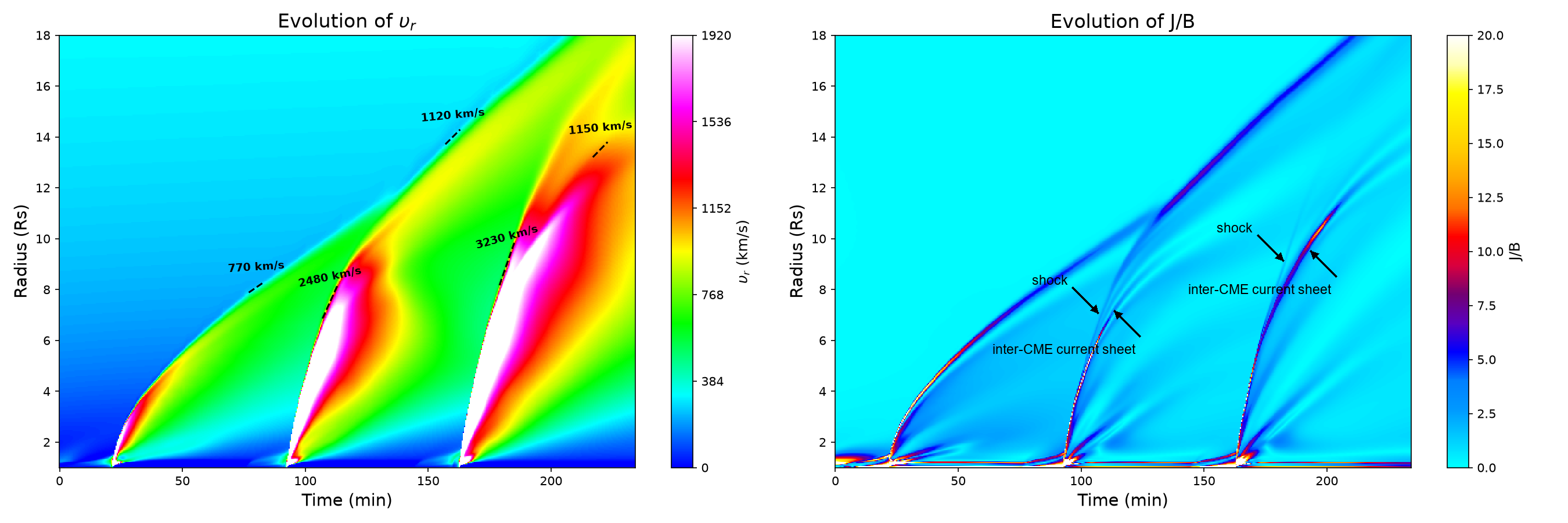}
    \caption{Time--distance maps of $v_r$ and $J/B$ along the eruption path indicated by the dashed line in Figure~\ref{fig:global}(b). In $v_r$ map, the typical speeds are denoted by the dashed lines and numbers. In $J/B$ map, the location of shocks and inter-CME current sheets are marked by the arrows.}
    \label{fig:td}
\end{figure*}

Figure~\ref{fig:td} shows time--distance maps of radial velocity and normalized current density along the eruption path marked in Figure~\ref{fig:global}b. As seen in the velocity map, all three CMEs complete their impulsive acceleration below $2\Rs$ and then undergo deceleration. At the same height of approximately $8\Rs$, the leading fronts, i.e., the shocks of the three CMEs, have markedly different speeds of approximately 770, 2480, and $3230\,\kms$, respectively. The larger speed of each later CME causes it to catch up with the preceding ejecta. Consequently, the CME2 shock catches the CME1 shock near $12\Rs$; after the shock fronts merge, the compound leading front propagates at approximately $1120\,\kms$. Because CME3 is faster than the compound ejecta formed by CME1 and CME2, it also catches up and begins to interact with that compound structure. The trajectories of the compound leading front and the CME3 front indicate that the CME3 system is merging with the preceding compound structure in the same manner as CME2 merges with CME1. Magnetic merging is expected to complete farther from the Sun, beyond the simulation domain.

\begin{figure*}
    \centering
    \includegraphics[width=\textwidth]{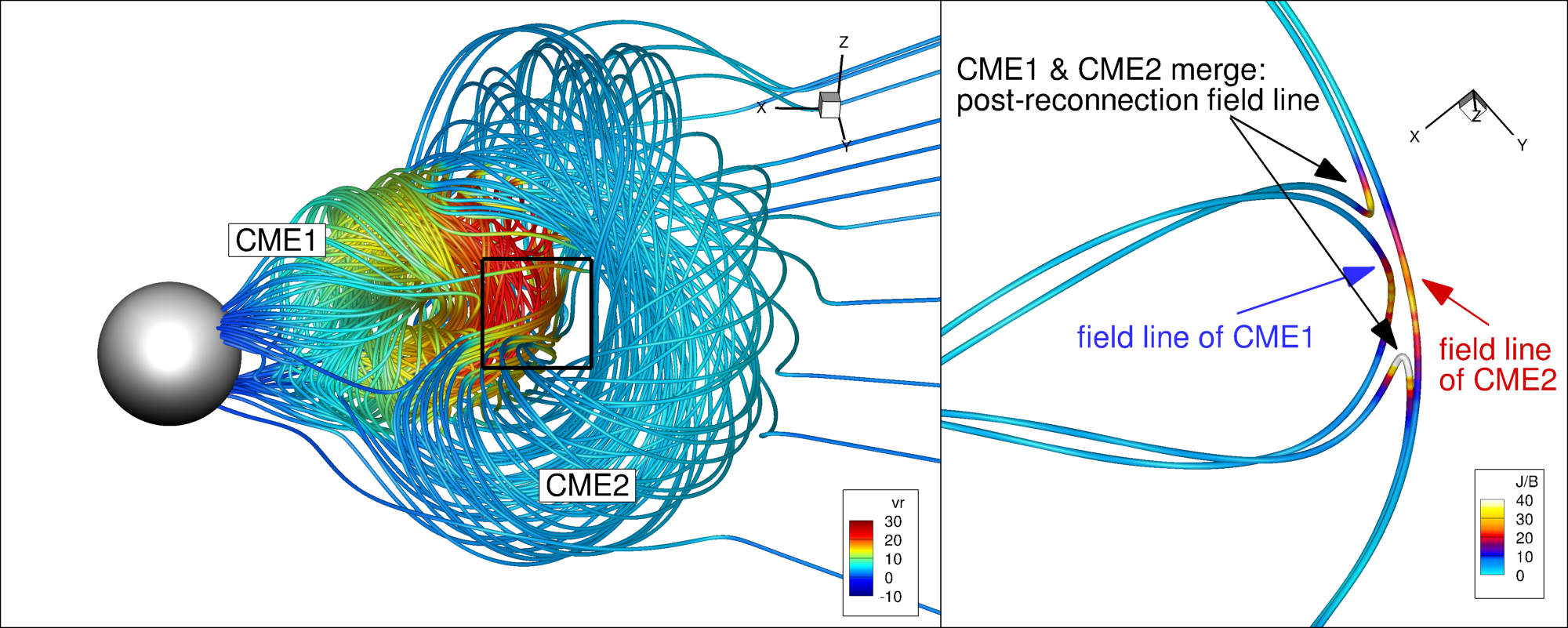}
    \caption{Magnetic reconnection between CME1 and CME2 sampled at $103.9$ minutes. The global view (left) shows field lines colored by normalized radial velocity (with unit of $128\,\kms$). The sphere represents solar surface. The boxed region is enlarged in the right panel with four sampled field lines illustrating the reconnection at the inter-CME current sheet.}
\label{fig:topology}
\end{figure*}

Besides showing the leading front of the CMEs, the $J/B$ map emphasizes evolution of the MFR boundaries. {After the onset of the second and third eruptions, the $J/B$ map initially shows a single front, which then bifurcates into two fronts, denoted by the two pairs of arrows in Figure~\ref{fig:td}. As mentioned in Section~\ref{subsec:global}, the outer front corresponds to the shock, across which the field lines are abruptly bent and the value of $J/B$ is therefore enhanced. The inner front, i.e., the MFR front, is a topological interface separating the newly erupted MFR from the preceding one. It constitutes a genuine current sheet formed between the MFRs of two consecutive CMEs, as can be elucidated by tracing field lines around this interface (see Figure~\ref{fig:topology}); it is referred to as the inter-CME current sheet.} Due to the homologous nature of the eruptions, the resulting CMEs have overall the same MFR topology before they interact with each other. Consequently, the magnetic field direction at the front of the later CME MFR is distinct from that in the wake of the preceding one, and when two MFRs interact, a current sheet naturally forms between them. It is the reconnection in this current sheet that merges the two CME magnetic flux systems.  

Figure~\ref{fig:topology} presents a snapshot of such reconnection between CME1 and CME2. At $103.9$~minutes, compression has brought oppositely directed components of the CME1 and CME2 fields together, forming the inter-CME current sheet. Field lines initially belonging to separate flux systems reconnect in this current sheet and form a shared twisted structure. As the CME2 magnetic front migrates through CME1 and reconnection changes the interface, the current sheet is gradually dissipated (see Figure~\ref{fig:global}b and its animation as well as Figure~\ref{fig:td}). By about $160$~minutes, the main bodies of the two CME MFRs have merged magnetically. Thus, shock catch-up, density-front overlap, and MFR merging occur at different times. The model therefore provides a magnetic interpretation for the evolution from multiple CME fronts to a compound ejecta as observed in coronagraph.

\subsection{Coronal Preconditioning}\label{subsec:precondition}

\begin{figure*}
    \centering
    \includegraphics[width=\textwidth]{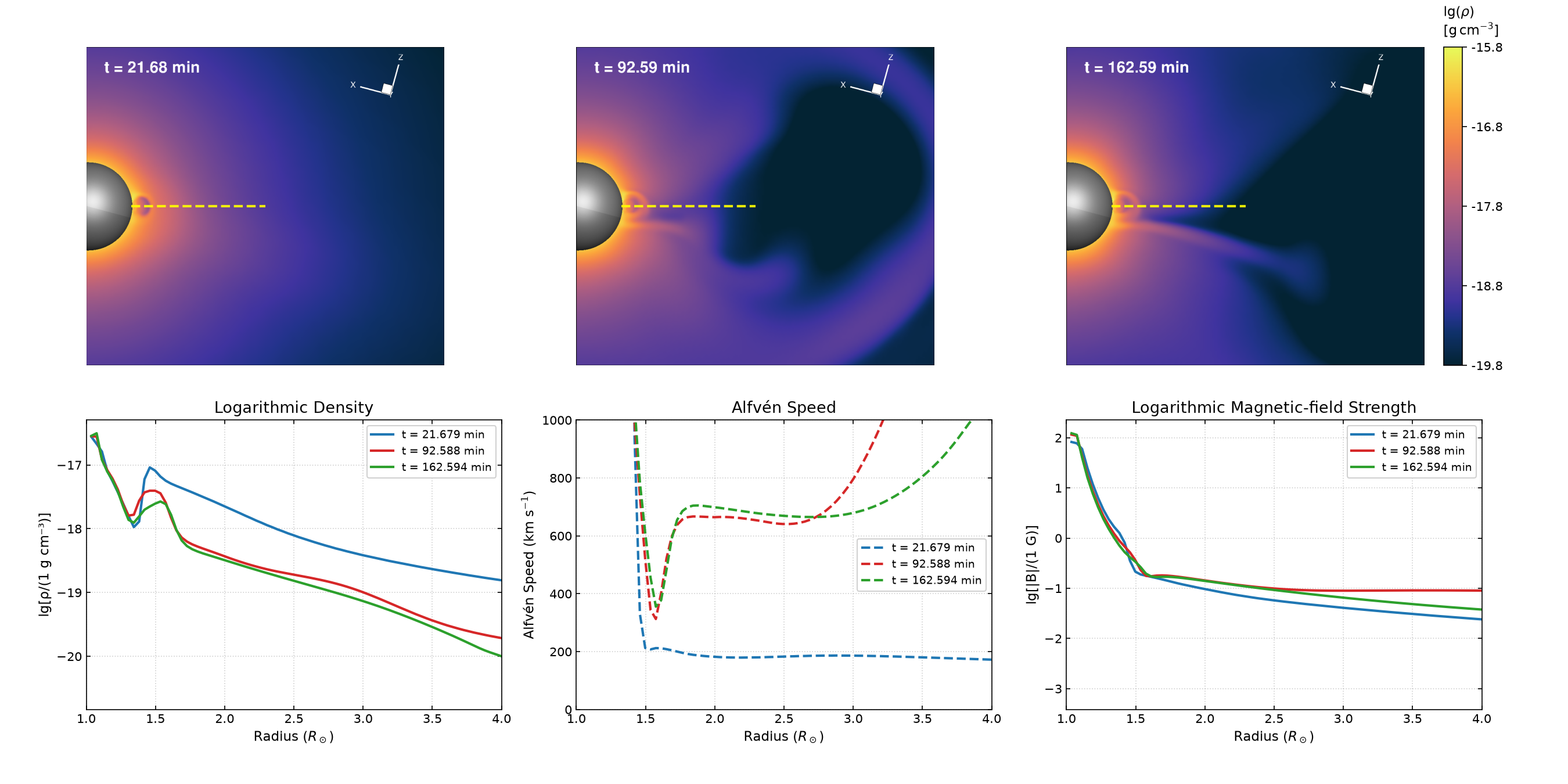}
    \caption{Coronal conditions immediately before the onset of three CMEs. Top: Density on the central cross section. Bottom: One-dimensional profiles of density, Alfv\'en speed, and magnetic field strength along the dashed path marked in the top panels.}
\label{fig:background}
\end{figure*}

The kinematics above show that CME2 and CME3 are substantially faster than CME1 and therefore catch up with the preceding ejecta. To identify why the later CMEs attain higher speeds, Figure~\ref{fig:background} compares the profiles of density, magnetic-field strength, and Alfv\'en speed at time immediately before the onset of three CMEs. The upper panels show the density on the central cross section. The lower panels show radial profiles sampled from solar surface to $4\Rs$ along the eruption path indicated by the dashed lines in the upper panels. This height range covers the overlying corona through which the CMEs complete their accelerations and therefore characterizes the environment relevant to CME acceleration.

Over much of the propagation path, the pre-CME2 and pre-CME3 densities are approximately one order of magnitude lower than the pre-CME1 density, demonstrating that the preceding eruptions progressively evacuate the overlying corona. The density profiles also exhibit progressively weaker enhancements associated with the initially rising MFRs (at around $1.5\Rs$), indicating that the mass loading of the erupting structures decreases from CME1 to CME3. These results, together with the larger kinetic-energy increments of CME2 and CME3 shown in Figure~\ref{fig:energy}d, explain the higher speeds of the later CMEs. Specifically, CME2 and CME3 receive more kinetic energy during eruption while carrying less plasma, and they subsequently propagate through a more rarefied corona in which less ambient material must be swept up. The combination of greater kinetic-energy release, lower ejecta mass loading, and reduced ambient inertia therefore produces the pronounced CME1--CME3 speed hierarchy.

The magnetic-field profiles further show that the field strength along part of the propagation path (beyond $1.5\Rs$) is enhanced before CME2 and CME3. Together with the substantial density depletion, this increases the local Alfv\'en speed, which rises from approximately $200\,\kms$  before CME1 to more than $600\,\kms$ along part of the preconditioned path before CME2 and CME3. The increased Alfv\'en speed corresponds to a higher local fast-mode characteristic speed. This elevated characteristic speed within the preceding ejecta helps the shocks of the later CMEs propagate rapidly and separate from their MFR fronts, as seen in Figure~\ref{fig:td}. The combined profiles thus identify reduced density as the dominant preconditioning effect, with the modified magnetic field and Alfv\'en speed governing shock transmission.

\section{Conclusions}\label{sec:conclusions}

For the first time, we present a global 3D MHD model of the initiation of homologous CMEs from a recurrently energized bipolar-arcade AR and their subsequent interactions. Starting from a potential field, the AR is supplied with free magnetic energy by three episodes of photospheric shearing flows, which result in three CMEs in sequence. In each episode, the AR coronal magnetic field evolves from an arcade to a sigmoid containing an internal current sheet, and an eruption is initiated by reconnection at the current sheet, ejecting a MFR into the outer corona while leaving a post-flare arcade in the AR.

Although the three eruptions have nearly identical initiation behavior in the AR, the resulting CMEs exhibit a systematic and significant increase in speed due to the progressive preconditioning of the corona by the preceding eruptions. That is, the later CMEs erupt into a more rarefied environment, with pre-eruption densities roughly an order of magnitude lower than before CME1. The faster later CME catches up with the former one, and the compression of the two magnetic flux systems forms an inter-CME current sheet. Then magnetic reconnection occurs at this current sheet, progressively merging the two CMEs into a compound ejecta. The synthetic coronagraph images also show the apparent coalescence of multiple coronagraph fronts into a single broad structure.  The simulation thus provides a magnetic explanation for the cannibalism of homologous CMEs. Our simulation also reproduces the propagation of the shock of the later CME into the preceding one before they fully merge. The Alfv\'en speed in the preconditioned corona increases by a factor of three, which explains why the shock front separates from the MFR front in CME2 and CME3, while the two fronts remain almost coincident in CME1.

The present simulation offers important insights into the complex dynamics of CME interactions. Future simulations covering the entire CME lifecycle from Sun to Earth will be performed to investigate how such interactions modulate key geoeffective parameters of the resulting ejecta, such as solar wind speed and southward interplanetary magnetic field strength at 1 au. More realistic simulations of homologous CME interactions will also be conducted using data-driven boundary conditions, enabling quantitative comparisons with observed events.

\vspace{1cm}
\noindent\textbf{Acknowledgments}: This work is jointly supported by the National Natural Science Foundation of China (NSFC 12573058), the National Key R\&D Program of China under No. 2024YFA1612001, the Shenzhen Science and Technology Program (grant No. RCJC20210609104422048), the Guangdong Basic and Applied Basic Research Foundation (2023B1515040021), the Fundamental Research Funds for the Central Universities
(HITJCRC2026005), and the Specialized Research Fund for the State Key Laboratory of Solar Activity and Space Weather.

\bibliography{references} 
\bibliographystyle{aasjournal} 

\end{document}